\documentclass[twocolumn,showpacs,prl,10pt]{revtex4}
\usepackage[english]{babel}
\usepackage{amsmath,amssymb} 
\usepackage{times}
\usepackage{epsfig} 

\begin{document} 
\title{Dynamical central peak and spinon deconfinement in frustrated
  spin chains}

\author{J. Kokalj$^1$ and P. Prelov\v sek$^{1,2}$}
\affiliation{$^1$J.\ Stefan Institute, SI-1000 Ljubljana, Slovenia}
\affiliation{$^2$ Faculty of Mathematics and Physics, University of
Ljubljana, SI-1000 Ljubljana, Slovenia}

\date{\today}
\begin{abstract}
Studying the dynamical spin structure factor in frustrated spin chains
with spontaneously dimerized ground state we show that besides the 
gapped spin-wave excitations there appears at finite temperatures
also a sharp central peak. The latter can be attributed to deconfined 
spinons, accounted well
within the variational approach. The central peak remains well
pronounced within the local spin dynamics and may be relevant 
for experiments on materials with 1D frustrated spin chains.
\end{abstract}

\pacs{71.27.+a, 75.10.Pq}

\maketitle

Frustrated spin systems have been intensively investigated both
theoretically an experimentally in last decades, offering novel
phenomena and challenges as well as a broader view on strongly
correlated electron systems. Among 1D models the spin-$1/2$
antiferromagnetic (AFM) Heisenberg chain (nearest-neighbor interaction
$J>0$) frustrated with the second neighbor AFM interaction $J'>0$
\cite{lecheminant} has attracted wide attention also due to its
relevance to the quasi-1D material CuGeO$_3$ exhibiting the
spin-Peierls transition at $T_{SP}=14$~K. For particular parameters
$J'/J=\alpha=0.5$ the exact ground state (g.s.) found by Majumdar and
Ghosh (MG) \cite{majumdar69} is doubly degenerate and dimerized with a
spin gap to excited states. Such a spin-liquid state without a long
range magnetic order has been shown to extend in a wider range around
this point, i.e. in the regime $\alpha>\alpha_c \sim
0.241$.\cite{okamoto92,eggert96,white96} Excited states at the MG
point have been determined analytically \cite{shastry81,caspers84} and
can be represented to a good approximation as a pairs of 
$S=1/2$ solitons or spinons with a gapped dispersion, the concept
confirmed by detailed numerical studies using the density-matrix
renormalization-group (DMRG) method.  \cite{sorensen98}

The dynamical properties of the frustrated $J$-$J'$ spin chain have
been so far mostly studied via the dynamical spin structure factor
$S(q,\omega)$ motivated again by the inelastic neutron scattering (INS) results on
CuGeO$_3$.\cite{arai96} At $T=0$ the continuum of $S=1$ excitation in
$S(q,\omega)$ in a wide range of $\alpha$ can be well represented with
a pair of spinons, \cite{yokoyama97,sorensen98} in particular if the phenomenologically
introduced matrix elements are taken into account \cite{singh96} in
analogy to the basic $\alpha=0$ AFM Heisenberg model.\cite{muller79} So far,
there are very few theoretical results on finite-$T$ properties of
frustrated system. It has been shown that the upper boundary of spinon
continuum in $S(q,\omega)$ persists even at $T>0$,\cite{fabricius98} while
the maximum in the static structure factor $S(q)$ exhibit a
shift to incommensurate $q<\pi$ at larger $\alpha$ and $T$. \cite{watanabe99,harada00}
 
 In the following we present evidence that $T>0$ dynamics of frustrated
spin-chain model with the spontaneously dimerized g.s. exhibits
several striking and rather unexpected features. Most
evident, numerically calculated $S(q,\omega)$ reveals at low but
finite $T>0$ a sharp central $\omega \sim 0$ peak well pronounced in
the region $q \sim \pi$, coexisting with the gapped two-spinon
continuum known already from $T=0$ studies.\cite{sorensen98, yokoyama97} The
central peak has its manifestation in a unusual $T$-dependence of
static susceptibility $\chi_{q \sim\pi}(T)$ with a maximum at
$T>0$. It shows up also in the local ($q$-integrated) $S_L(\omega)$ as
relevant, e.g., for the NMR spin-lattice relaxation. Using a variational presentation of
excited two-spinon states and relevant matrix elements we show that the
phenomenon can be directly traced back to spinons and their deconfined
nature.
 
In the following we study the frustrated spin model on a 1D chain
\begin{equation}
H=J \sum_i [{\bf S}_i \cdot {\bf S}_{i+1}+\alpha {\bf S}_i \cdot
{\bf S}_{i+2}], \label{ham}
\end{equation}
where ${\bf S}_i$ are local $S=1/2$ operators and the only relevant
parameter is $\alpha=J'/J$ (we choose furtheron $J=1$). The model has
been invoked as the microscopic model for CuGeO$_3$ with $\alpha \sim
0.36$ (realized above $T<T_{SP}$ where lattice-deformation-induced
dimerization is zero). But it as well represents the 1D zig-zag spin system,
example being double-chain compound SrCuO$_2$ within the opposite
limit of large $|\alpha| \sim 5 - 10$. \cite{matsuda97}

As the central quantity we calculate dynamical $S(q,\omega)$ at $T>0$.
We employ two numerical approaches. Finite-$T$ Lanczos method (FTLM)
\cite{jaklic94} based on the Lanczos diagonalization of small systems
covers the whole $T$ range but is restricted to system sizes $N \leq 28$
whereby we use periodic boundary conditions (b.c.).  Finite-$T$
dynamical extension of the DMRG (FTD-DMRG) method recently developed
by the present authors \cite{kokalj09} combines the DMRG optimization
of basis states with the FTLM method for dynamical correlations at
$T>0$ and offers more powerful method for low $T$. The model,
Eq.(\ref{ham}), is here studied with open b.c. The reachable system
sizes depend on $T$ and the method shows good convergence, at least
for low $\omega$, for systems
$N<60$ for $T \leq 0.5$ presented here, with the typical subblock
dimension $m \leq 256$. Concentrating on the low-$\omega$ dynamical
window the method is used as presented in Ref.\cite{kokalj09}, while
high-$\omega$ results are improved by the application of correction
vectors increasing at the same time computation demand.  The advantage
of both methods is very good spectral resolution, so that typically
only a minor additional $\omega$ dependent broadening of $\delta \sim
0.02$ at $\omega \sim 0$ and $\delta \sim 0.06$ at higher $\omega$ is
employed in presentations.

Let us first present results for the MG model with $\alpha=0.5$. While
$S(q,\omega)$ at $T=0$ is rather well understood and investigated
numerically,\cite{yokoyama97} we concentrate in Figs.~1 and 2 on $T>0$ FTD-DMRG
results for different $T/J \leq 0.5 $. The high-$\omega$ continuum
appearing at $T=0$ above the two-spinon gap $\Delta_0 \sim 0.25 J$
\cite{white96} is qualitatively not changed from $T=0$ spectra. The
evident new feature is the central peak at $\omega \sim 0$ most
pronounced at $q \sim \pi$. Its width $w$ is very narrow but still of
intrinsic nature (being larger than the additional broadening $\delta
=0.02$). It is evident that at fixed $T$ the width $w$ increases away from
$q =\pi$ whereby the peak also looses the intensity. Still it remains
well pronounced in wide region $q>0.7 \pi$. The comparison of Figs.~1
and 2
also reveals that $w$ increases as well with $T$ and finally
merges in a broader continuum for high $T$.

\begin{figure}[htb]
\includegraphics[angle=-90, width=1.\linewidth]{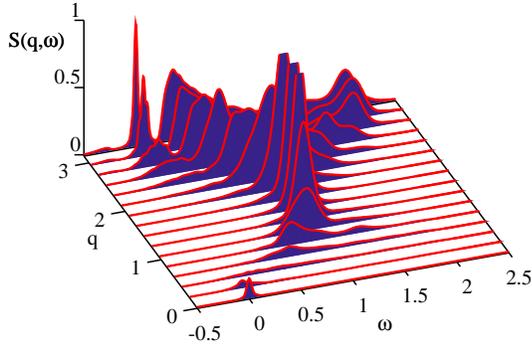}
\caption{(Color online) Dynamical spin structure factor
$S(q,\omega)$ for the MG model at $T/J=0.1$ within the whole range 
of  $q \leq \pi$, calculated by the FTD-DMRG method on a system of
  $N=60$ sites. Correction vector improvement of high $\omega$ part was
preformed only for $q \geq 13\pi/15$.}
\label{fig1}
\end{figure}
 
\begin{figure}[htb]
\includegraphics[angle=-90, width=.9\linewidth]{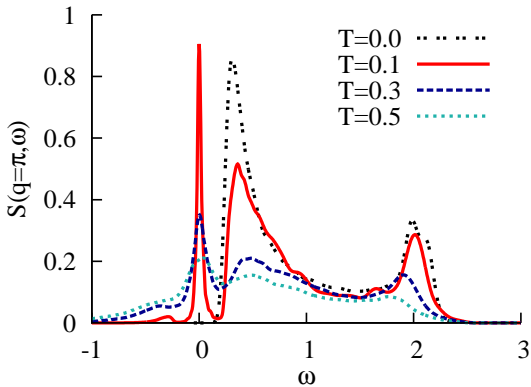}
\caption{(Color online) $S(q=\pi,\omega)$ for different $T/J=0.0, 0.1, 0.3,
  0.5$ where the broadening of central peak with increasing $T$ is well pronounced.}
\label{fig2}
\end{figure}

In the following we present the analysis showing that the emergence of the
central peak in $S(q,\omega)$ at low $T<J$ can be described well in
terms of spinons as relevant excitations of the system and their
deconfinement. At the MG point $\alpha=0.5$ the g.s. (for even $N$ and
periodic b.c.) has the energy $E_0=-3NJ/8$ and the wavefunction which
can be written as the product of local singlets
$\Psi_0=[1,2][3,4]\cdot[N-1,N]$. It is doubly degenerate with the
corresponding eigenstate $\tilde \Psi_0$ having for one site shifted
singlets. It has been already realized \cite{shastry81, caspers84,
  lecheminant} that lowest excitations can be well represented in
terms of spinon states. In particular, the lowest branch of
approximate triplet ($S=1,S^z=1$) eigenstates can be constructed from
the local triplet two-spinon states
\begin{eqnarray}
&\psi^t(p,m)=[1,2]\ldots [2p-3,2p-2]\uparrow_{2p-1}[2p,2p+1]\ldots \nonumber\\
&[2m-2,2m-1] \uparrow_{2m}[2m+1,2m+2]\ldots. \label{psipm}
\end{eqnarray}
where the first spinon is on site $2p-1$ and the second on site $2m$.
Since the total momentum $Q$ is conserved due to periodic b.c., the
relevant two-spinon functions are
\begin{equation}
\psi^t_Q(k)=\frac{1}{M}\sum_{p,m}^M e^{i(Q+k)p+i(Q-k)m} \psi^t(p,m),
\label{psiqk}
\end{equation}
where sums run over $M=N/2$ double cells. In the further analysis
difficulties arise since $\psi^t(p,m)$ are not orthogonal even for
distant $|p-m| \gg 1$ and furthermore for $p \sim m$. To find proper
eigenfunctions we follow the procedure and notation of Ref.\cite{shastry81}
which for each momentum subspace $Q$ yields nontrivial matrix elements
\begin{eqnarray}
\langle \psi^t_{Q}(k')|\psi^t_{Q}(k)\rangle&=&\frac{9 J^2}
 {64 \omega_-\omega_+}\delta_{k,k'}+\frac{1}{M}\chi_Q(k,k'),  \label{psikk} \\
\langle \psi^t_{Q}(k')|\tilde H|\psi^t_{Q}(k)\rangle&=&\frac{9
\epsilon_Q(k)J^2} {64 \omega_-\omega_+ }\delta_{k,k'}+\frac{1}{M}h_Q(k,k'),
\nonumber
\end{eqnarray}
where $\tilde H=H-E_0$, $\omega_\pm=\omega((Q\pm k)/2)$,
$\omega(p)=(5/4+\cos 2p) J/2$ are (approximate) single-spinon energies
and
\begin{equation}
\epsilon_Q(k)=\omega_+ +\omega_-= (\frac{5}{4}+\cos Q \cos k)J.
\label{e2spin}
\end{equation}
The off-diagonal terms $\chi_Q(k,k'),h_Q(k,k')$ (not presented here)
emerging from nonorthogonality of $\psi^t_Q(k)$ are the same as given
in Ref.\cite{shastry81}. Within the triplet two-spinon basis,
Eqs.~(\ref{psipm}) and (\ref{psiqk}), proper eigenstates (nevertheless
not yet exact eigenstates of Eq.~(\ref{ham})) are obtained via the
diagonalization of Eqs.(\ref{psikk}) and can be denoted $\Psi^t_Q(k)$
(whereby $k$ remains only a label and not a well defined
wavevector). In contrast to $\psi^t_Q(k)$, $\Psi^t_Q(k)$ are
ortho-normalized. The corresponding two-spinon excitation energies 
$e^t_Q(k)$ are well approximated as the sum of two (deconfined) free
spinons, i.e. $e^t_Q(k) \sim \epsilon_Q(k)$, Eq.(\ref{e2spin}). 

Our goal is, however, to understand low-$T$ properties of
$S(q,\omega)$. Before discussing $T>0$ results, we first have to
reconsider the $T=0$ spectrum $S_0(q,\omega)$ which has been already
interpreted in terms of two-spinon excitations.\cite{poilblanc97,
  sorensen98, yokoyama97}
Still the corresponding matrix element has been postulated so far
only phenomenologically \cite{singh96,shastry97} in analogy with previous
works on the unfrustrated Heisenberg model.\cite{muller79} We note that at
$T=0$ within the chosen subspace, Eq.(\ref{psipm}), we can express
\begin{equation}
S_0(q,\omega)=\frac{1}{2}\sum_k |\langle \Psi_q^t(k)|S^+_q|\Psi_0
\rangle|^2 \delta(\omega-e_q(k)), \label{sq0om}
\end{equation}
and 
\begin{equation}
S^+_q|\Psi_0\rangle
= \frac{1-e^{-iq}}{2 M} \sum_k
\psi^t_q(k). \label{sqop}
\end{equation}
Since Eq.(\ref{sqop}) is an exact representation of the $S^+_q$
operator, we can evaluate matrix elements
$\zeta_q(k)=\langle\Psi^t_q(k)|S^+_q|\Psi_0 \rangle$ by
diagonalizing numerically equations, Eqs.(\ref{psikk}). Taking into
account that $e^t_q(k) \sim \epsilon_q(k)$ we can then evaluate
$S_0(q,\omega)$ in the two-spinon approximation. Results within such a
framework are presented for $q=\pi$ in Fig.~3 along with the full
numerical results obtained via the $T=0$ FTD-DMRG (for $T=0$ identical
to the more standard dynamical DMRG) evaluated within a system of $N=100$
sites. The agreement is very satisfactory except at the higher-$\omega$ end
where the obtained intensity is too low as well as Eq.(\ref{psikk})
seem to generate a high two-spinon anti-bound state (peak in
$S_0(\pi,\omega)$) besides the free two-spinon dispersion,
Eq.(\ref{e2spin}). It should be noted that obtained $\zeta_q(k)$ is
quite far from the oversimplified spinon picture with $\zeta_q(k)
\sim 1$.\cite{shastry97} Still it is hard to find for it an
appropriate analytical expression.\cite{shastry81,shastry97} One possibility is
to neglect non-orthogonalities in Eqs.(\ref{psikk}) which yields $\tilde \zeta_q(k)
\propto 1/(\omega_+\omega_-)^{1/2}$.  Corresponding ``free" spinons
results for $S_0(q=\pi,\omega)$ also presented in Fig.~3 show
qualitatively reasonable trend (fall-off for higher $\omega$). Still
they give an incorrect behavior at lower and higher cut-off due to divergent
two-spinon density of states.

\begin{figure}[htb]
\includegraphics[angle=-90, width=.9\linewidth]{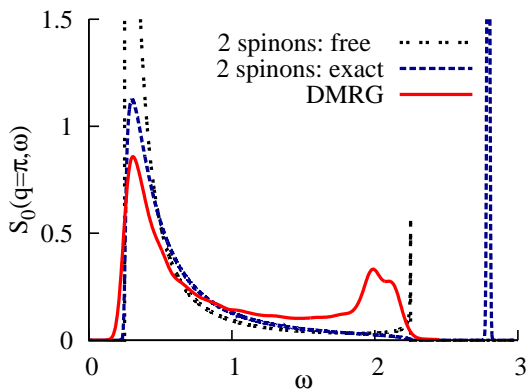}
\caption{(Color online) $T=0$ dynamical spin structure factor
  $S_0(q=\pi,\omega)$ as calculated via the DMRG for $N=100$ sites (full
  line), numerically within the two-spinon approximation (dashed line)
  and using simplified $\tilde \zeta_q(k)$ (dotted line).}
\label{fig3}
\end{figure}

The above agreement of numerical $T=0$ results with the description in
terms of the two-spinon basis, Eq.(\ref{psipm}), gives firm support
also to the interpretation of $T>0$ dynamics. In the low-$T$ regime we
are in $S(q,\omega)$ predominantly dealing with excitations increasing
the number of spinons, $n_s \to n_s + 2$, analogous to those in
$S_0(q,\omega)$, Eqs.~(\ref{sq0om}) and (\ref{sqop}).  Their contribution
analogous to $T=0$ Fig.~3 is evident also at $T>0$ in
Figs.~1 and 2. 

However, in addition there are possible transitions between excited
states conserving $n_s$. In particular, the matrix element
$\gamma_{qQ}(k,k')=\langle \Psi^t_{q+Q}(k)|S^+_q|\Psi^s_Q(k')\rangle$
as introduced already in Ref.\cite{shastry97} is finite and
nontrivial.  Here $\Psi^s_Q(k')$ are singlet two-spinons eigenstates.
We evaluate $\gamma_{qQ}(k,k')$ numerically assuming two-spinon
approximation, Eq.({\ref{psikk}). Results show that elements are
  nearly diagonal, i.e., $\gamma_{qQ}(k,k') \sim \delta_{k',k+Q}$
  leading in $S(q,\omega)$ to the contribution at $\omega \sim
  e^t_{q+Q}(k+q)-e^s_Q(k)$. Since at low-$T$ favored are lowest
  excited states, i.e., from Eq.(\ref{e2spin}) $Q \sim 0,k \sim \pi$
  and $Q \sim \pi,k \sim 0$ with a Boltzmann weight $p \propto \exp(-
  \Delta_0/T)$ (where $\Delta_0\sim\epsilon_{\pi}(0)=J/4$). Numerical
  solution of two-spinon problem shows that $e^s_{Q}(k) \sim
  e^t_{Q}(k) \sim \epsilon_Q(k)$ consistent with the picture of
  unbound (deconfined) spinons. Hence, the strongest transitions are
  at $\omega \sim \epsilon_{q+Q}(k+q)-\epsilon_Q(k)$.  This evidently
  leads at $q \sim \pi$ to a sharp central peak at $\omega \sim 0$
  with the strength increasing as $\propto \exp(-\Delta_0/T)$.
  
From above perspective we therefore conclude that the pronounced central peak in
Figs.~1,2 at $q \sim \pi$ confirm the presented analysis of nearly-free
or deconfined spinons as excited states at the MG point. On the other hand, 
even without the extensive calculations it is evident that the central
peak can only appear if triplet and singlet spinon states are nearly degenerate
again only possible for deconfined spinons. 

The emergence of the central peak is, however, not restricted to the
MG point $\alpha=0.5$ but appears to be related closely to the
existence of the spontaneous dimerization at $\alpha>\alpha_c$ and the
spin gap $\Delta_0>0$. We tested numerically also the case
$\alpha=0.7$ (only partly presented here) where the spin gap is larger
$\Delta_0 \sim 0.4 J$. \cite{white96} Consequently also the central
peak feature is even more pronounced and extended in the $q$ space
as well as persists to higher $T$.

It is evident that the central peak has a substantial effect on
the static susceptibility $\chi_q(T)$,
\begin{equation}
\chi_q(T)= \int_{-\infty}^{\infty} \frac{d\omega}{\omega} [1-
  {\rm e}^{-\omega/T}] S(q,\omega), \label{chiq}
\end{equation}
being sensitive to low-$\omega$ dynamics.  In Fig.~3 we show the
FTLM results obtained on systems with $N=28$ sites for $\chi_q(T)$
with various $q$ and again $\alpha = 0.5$.  Most pronounced is
the variation at $q=\pi$ where the g.s. value $\chi_\pi(0)$ is
determined with the dimerization gap, i.e. $\chi_\pi(0) \propto
1/\Delta_0$. Instead of naively expected monotonously decreasing
$\chi_q(T)$, we observe in Fig.~\ref{fig4} simultaneously with the emergence of
the central peak an increasing $\chi_\pi(T)$ in the regime $0<T<T^*$
whereby $T^* \sim \Delta_0/2$. On the other hand, for $T>T^*$ the
fall-off is uniform with $\chi_\pi(T) \propto 1/T$ which is close to
the characteristic critical $q=\pi$ behavior for the simple AFM
Heisenberg model.\cite{kokalj09} Results for $q< \pi$ are quite analogous
taking into account that the relevant spin gap is $\Delta_q >
\Delta_0$.

\begin{figure}[htb]
\includegraphics[angle=-90, width=.9\linewidth]{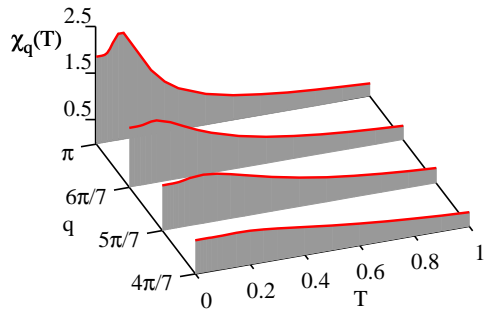}
\caption{(Color online) Static susceptibility $\chi_q(T)$ vs. $T$ for
  different $q\leq \pi$ for $\alpha =0.5$ as obtained with
  the FTLM.}
\label{fig4}
\end{figure}

The effect of the central peak is visible also in local spin
correlations $S_L(\omega)=(1/N)\sum_q S(q,\omega)$ as presented in
Fig.~\ref{fig5}. They are accessible directly via FTD-DMFT by
calculating local spin correlations $\langle S^z_i, S^z_i
\rangle_\omega$ locating the site $i \sim N/2$ to avoid effects of
open b.c. As well we can evaluate them within the FTLM and periodic
b.c. summing all $S(q \neq 0,\omega)$ ($q=0$ contribution is delta
function due to the conserved $S_{tot}^z$).  Clearly, in both
approaches the diffusion contribution $q \sim 0$ is not represented
correctly but it is expected to be subdominant.\cite{itoh96} FTD-DMRG
results in Fig.~5 presented for $\alpha=0.5, 0.7$ reveal a central peak
at $\omega \sim 0$ well separated from the higher-$\omega$ two-spinons
continuum as far as $T \lesssim \Delta_0$. The peak gains the weight at
$T \sim T^*$ and for $T>T^*$ steadily becomes broader,
finally merging with the continuum for $T> \Delta_0$. Moreover we
observe for both $\alpha$ that $S_L(\omega=0)$ is nearly constant in a
broad range $T^* <T< 2J$.

\begin{figure}[htb]
\includegraphics[angle=-90, width=.8\linewidth]{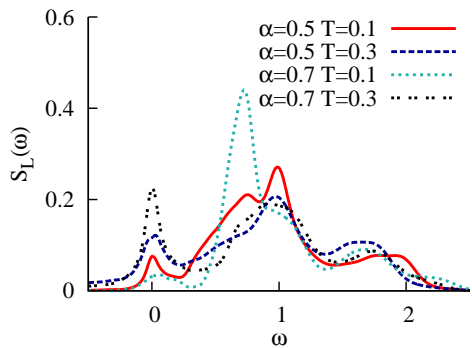}
\caption{(Color online) Local spin correlations $S_L(\omega)$ for
  $\alpha=0.5, 0.7$ at $T=0.1, 0.3$ as obtained with the
  FTD-DMRG method.}
\label{fig5}
\end{figure}

It should be noted that within the simplest approximation (with
$q$-independent form factor) the NMR or NQR spin-lattice relaxation
should be a closely related to $S_L(\omega)$, i.e. the relaxation rate
is given by $1/T_1 \propto S_L(\omega=0)$. Following above results we
would obtain for considered systems $1/T_1 \sim$~const in a
broad range $T>T^*$ similar to theoretical predictions for the 1D
(unfrustrated) AFM Heisenberg model and CuGeO$_3$ \cite{itoh96}. While the
agreement for higher $T>\Delta_0$ with the Heisenberg model is not
surprising the novel contribution of the central peak is that the
validity of this universality is extended to lower $T>T^*$. It should
be also reminded that such relaxation is far from the usual Korringa
relaxation with $1/(TT_1) \sim$~const.

In conclusion, we have shown that the frustrated spin chain as manifested within 
the 1D $J$-$J'$ model with $\alpha>\alpha_c$ reveals besides the gap in spin excitations at 
$T=0$ also very unusual spin dynamics at finite but low $T<\Delta_0$. The central peak 
which appears in $S(q,\omega)$ at $q\sim \pi$ as well in the $q$-integrated 
local $S_L(\omega)$ is very sharp and dominates the low-$\omega$ response 
at low $T$. It is a direct consequence and the signature of deconfinement of spinon excitations
in such systems. It remains to be investigated whether such a behavior 
is restricted to the particular case of investigated model or there are other gapped spin
systems with similar phenomena. As far as experimental relevance is concerned 
extensively investigated CuGeO$_3$ above the spin-Peierls transition $T>T_{SP}$ is
interpreted with a frustrated spin-chain model with $\alpha \sim 0.36$
and could partly exhibit  
mentioned phenomena in spite of presumably very small scale $\Delta_0<0.02 J$
\cite{sorensen98}.    
 
We authors acknowledge helpful discussions with T. Tohyama 
as well as the support of the Slovenia-Japan Research Cooperative
grant and the Slovenian Agency grant No. P1-0044.


\begin{thebibliography}{21}

\bibitem{lecheminant}
for a review~see P.~Lecheminant, \emph{Frustrated Spin Systems} (ed. H. T.
  Diep, World Scientific, p.307, 2004).

\bibitem{majumdar69}
C.K. Majumdar, D.K. Ghosh, J. Math. Phys. \textbf{10}, 1388, 1399 (1969).

\bibitem{okamoto92}
K.~Okamoto, K.~Nomura, Phys. Lett. A \textbf{169}, 433  (1992).

\bibitem{eggert96}
S.~Eggert, Phys. Rev. B \textbf{54}, R9612 (1996).

\bibitem{white96}
S.R. White, I.~Affleck, Phys. Rev. B \textbf{54}, 9862 (1996).

\bibitem{shastry81}
B.S. Shastry, B.~Sutherland, Phys. Rev. Lett. \textbf{47}, 964 (1981).

\bibitem{caspers84}
W.J. Caspers, K.M. Emmett, W.~Magnus, J. Phys. A \textbf{17}, 2687 (1984).

\bibitem{sorensen98}
E.~S\o{}rensen, I.~Affleck, D.~Augier, D.~Poilblanc, Phys. Rev. B \textbf{58},
  R14701 (1998).

\bibitem{arai96}
M.~Arai, M.~Fujita, M.~Motokawa, J.~Akimitsu, S.M. Bennington, Phys. Rev. Lett.
  \textbf{77}, 3649 (1996).

\bibitem{yokoyama97}
H.~Yokoyama, Y.~Saiga, J. Phys. Soc. Jpn. \textbf{66}, 3617 (1997).

\bibitem{singh96}
R.R.P. Singh, P.~Prelov\ifmmode~\check{s}\else \v{s}\fi{}ek, B.S. Shastry,
  Phys. Rev. Lett. \textbf{77}, 4086 (1996).

\bibitem{muller79}
G.~M\"uller, H.~Beck, J.C. Bonner, Phys. Rev. Lett. \textbf{43}, 75 (1979).

\bibitem{fabricius98}
K.~Fabricius, U.~L\"ow, Phys. Rev. B \textbf{57}, 13371 (1998).

\bibitem{watanabe99}
S.~Watanabe, H.~Yokoyama, J. Phys. Soc. Jpn \textbf{68}, 2073 (1999).

\bibitem{harada00}
I.~Harada, Y.~Nishiyama, Y.~Aoyama, S.~Mori, J. Phys. Soc. Jpn, Suppl. A
  \textbf{69}, 339 (2000).

\bibitem{matsuda97}
M.~Matsuda, K.~Katsumata, K.M. Kojima, M.~Larkin, G.M. Luke, J.~Merrin,
  B.~Nachumi, Y.J. Uemura, H.~Eisaki, N.~Motoyama et~al., Phys. Rev. B
  \textbf{55}, R11953 (1997).

\bibitem{jaklic94}
J.~Jakli\ifmmode~\check{c}\else \v{c}\fi{}, P.~Prelov\ifmmode~\check{s}\else
  \v{s}\fi{}ek, Phys. Rev. B \textbf{49}, 5065 (1994).

\bibitem{kokalj09}
J.~Kokalj, P.~Prelov\ifmmode~\check{s}\else \v{s}\fi{}ek, Phys. Rev. B
  \textbf{80}, 205117 (2009).

\bibitem{poilblanc97}
D.~Poilblanc, J.~Riera, C.A. Hayward, C.~Berthier, M.~Horvati\'c, Phys. Rev. B
  \textbf{55}, R11941 (1997).

\bibitem{shastry97}
B.S. Shastry, D.~Sen, Phys. Rev. B \textbf{55}, 2988 (1997).

\bibitem{itoh96}
M.~Itoh, M.~Sugahara, T.~Yamauchi, Y.~Ueda, Phys. Rev. B \textbf{54}, R9631
  (1996).

\end{thebibliography}

\end{document}